\documentclass[aps, prl, twocolumn, notitlepage, superscriptaddress]{revtex4-1}
\usepackage[T1]{fontenc}
\usepackage{amsmath}
\usepackage{hyperref}
\usepackage{booktabs}
\usepackage{appendix}
\usepackage[inline]{enumitem}
\usepackage{multirow}
\usepackage{amssymb}
\usepackage{amsthm}
\usepackage{array}
\usepackage{graphicx}
\usepackage{verbatim}
\usepackage{bbm}
\usepackage{color}
\usepackage[normalem]{ulem}
\DeclareGraphicsExtensions{.png,.pdf,.eps}

\def \diracspacing {0.7pt}
\newcommand{\ket}[1]{| \hspace{\diracspacing} #1 \rangle} 
\newcommand{\ketbra}[2]{| \hspace{\diracspacing} #1 \rangle \langle #2 \hspace{\diracspacing} |} 
\newcommand{\ketbraq}[1]{\ketbra{#1}{#1}} 
\newcommand{\prlparagraph}[1]{\emph{#1.---}}
\newcommand{\SupMat}{See the Supplemental Material.}

\newcommand{\I}{\mathbb{I}}

\newcommand{\cL}{\mathcal{L}}
\newcommand{\cNL}{\mathcal{N\!L}}
\newcommand{\cH}{\mathcal{H}}

\newcommand{\va}{\vec{\alpha}}
\newcommand{\vb}{\vec{\beta}}
\newcommand{\la}{\langle}
\newcommand{\ra}{\rangle}

\newcommand{\rt}{r}

\DeclareMathOperator{\tr}{tr}

\theoremstyle{definition}
\theoremstyle{plain}

\theoremstyle{remark}

\begin{document}
\title{Bell nonlocality is not sufficient for the security of standard device-independent quantum key distribution protocols}
\author{M\'at\'e Farkas}
\email{mate.farkas@icfo.eu}
\affiliation{ICFO -- Institut de Ciencies Fotoniques, The Barcelona Institute of Science and Technology, 08860 Castelldefels, Spain}
\author{Maria Balanz\'o-Juand\'o}
\affiliation{ICFO -- Institut de Ciencies Fotoniques, The Barcelona Institute of Science and Technology, 08860 Castelldefels, Spain}
\author{Karol \L ukanowski}
\affiliation{Centre for Quantum Optical Technologies, Centre of New Technologies, University of Warsaw, Banacha 2c, 02-097 Warszawa, Poland}
\affiliation{Faculty of Physics, University of Warsaw, Pasteura 5, 02-093 Warszawa, Poland}
\author{Jan Ko\l ody\'nski}
\affiliation{Centre for Quantum Optical Technologies, Centre of New Technologies, University of Warsaw, Banacha 2c, 02-097 Warszawa, Poland}
\author{Antonio Ac\'in}
\affiliation{ICFO -- Institut de Ciencies Fotoniques, The Barcelona Institute of Science and Technology, 08860 Castelldefels, Spain}
\affiliation{ICREA-Instituci\'o Catalana de Recerca i Estudis Avan\c{c}ats, Lluis Companys 23, 08010 Barcelona, Spain}

\begin{abstract}
Device-independent quantum key distribution is a secure quantum cryptographic paradigm that allows two honest users to establish a secret key, while putting minimal trust in their devices. Most of the existing protocols have the following structure: first, a bipartite nonlocal quantum state is distributed between the honest users, who perform local measurements to establish nonlocal correlations. Then, they announce the implemented measurements and extract a secure key by post-processing their measurement outcomes.
We show that no protocol of this form allows for establishing a secret key when implemented on any correlation obtained by measuring local projective measurements on certain entangled nonlocal states, namely on a range of entangled two-qubit Werner states. To prove this result, we introduce a technique for upper-bounding the asymptotic key rate of device-independent quantum key distribution protocols, based on a simple eavesdropping attack. Our results imply that either different reconciliation techniques are needed for device-independent quantum key distribution in the large-noise regime, or Bell nonlocality is not sufficient for this task.
\end{abstract}

\maketitle

\prlparagraph{Introduction}
Device-independent quantum key distribution (DIQKD) is the strongest form of quantum cryptographic protocols \cite{ABG+07,PAB+09}. DIQKD security proofs are based on the assumption that quantum theory is correct, and on physically observable quantities. Most importantly, the honest users do not need to make any assumptions about the inner workings of their devices and, hence, they do not need to trust the preparation by the manufacturer. These facts make DIQKD a promising paradigm for providing complete security, in that DIQKD protocols are not vulnerable to implementation flaws that may be exploited in hacking attacks~\cite{ZFQ+08,LWW+10,GLL+11}.

In DIQKD protocols, two distant honest parties aim at sharing a cryptographic key, while assuring it to be unknown to any eavesdropper limited by quantum theory. To achieve this, a quantum state is distributed between them in each round of the protocol, and in each of these rounds they measure the part of the state that is available to them. The resulting set of measurement outcomes is the raw data from which they extract the secure key. This data
is characterised by the set of measurement outcome probabilities, called the \textit{correlation}. If the observed correlation violates a Bell inequality, the information of any quantum eavesdropper about the outcomes is limited, which opens up the possibility of extracting a secure key.

The first quantitative security proofs were achieved based on the violation of the Clauser--Horne--Shimony--Holt (CHSH) Bell inequality \cite{CHSH69}, using one-way public communication for key extraction \cite{ABG+07,PAB+09}. The original proofs, valid under collective attacks, were later extended to the most
powerful coherent attacks~\cite{VV14,Arn20,ADF+18}. Furthermore, moderate improvements on key rates were recently achieved using the more general \textit{biased} CHSH inequalities~\cite{AMP12}, and applying noisy pre-processing~\cite{WAP20,HST+20,SBV+20}. Similarly, advantage distillation---a protocol for key distillation using two-way public communication---has been shown to be useful under the assumption of collective attacks \cite{TLR20}.

While nonlocality of the observed correlations is necessary for DIQKD, it remains an open question whether it is also sufficient. Recently, methods have been proposed for upper-bounding secure key rates in DIQKD, even if two-way communication is allowed \cite{KWW20,CFH20,AL20}. However, all the bounds constructed thus far remain strictly positive for nonlocal correlations, suggesting that nonlocality may be sufficient for DIQKD.

In this work, we show this not to be the case for DIQKD protocols consisting of the following two steps: (i)~nonlocal correlations are established by applying local measurements on an entangled quantum state; (ii)~the implemented measurements are announced and the key is constructed by classically post-processing the outcomes. Most of the existing DIQKD protocols have this form and, hence, in what follows we refer to such protocols as \emph{standard} protocols. To prove the result, we provide a generic tool for upper-bounding key rates in DIQKD. We then apply our tool to standard protocols implemented on a two-qubit Werner state \cite{Wer89} using an arbitrary number of projective measurements. We show that for a range of visibilities for which the Werner state is known to be nonlocal, the upper bound on the key rate is zero, and therefore no standard DIQKD protocol can be secure. This means that there exist nonlocal correlations that cannot be used for standard DIQKD, and, furthermore, that there exist nonlocal quantum states that cannot be used for standard DIQKD with projective measurements. We also show how the provable region of insecurity can be enlarged when fixing the number of measurements in the protocol.  In particular, we compute visibilities for which the commonly used protocols based on the (biased) CHSH inequality \cite{ABG+07,PAB+09,ADF+18,WAP20,HST+20,SBV+20,TLR20} all become insecure despite the correlations still being nonlocal.


\prlparagraph{Methods}
Formally, in a DIQKD protocol two parties, Alice and Bob, have access to a bipartite quantum state, $\rho_{AB}$, represented by a positive semidefinite operator with unit trace on the tensor product Hilbert space, $\cH_A \otimes \cH_B$. The protocol consists of several rounds, in each of which Alice and Bob choose a particular quantum measurement to measure their part of a fresh copy of $\rho_{AB}$. In particular, Alice chooses a measurement labelled by $x \in \{0, 1, \ldots, n_A - 1\} \equiv [n_A]$, and Bob chooses a measurement labelled by $y \in [n_B]$. Without loss of generality, we assume that each of Alice's (Bob's) measurements has $k_A$ ($k_B$) possible outcomes. According to quantum theory, $k_A$-($k_B$-)outcome measurements correspond to a set of $k_A$ ($k_B$) positive semidefinite operators on $\cH_A$ ($\cH_B$), adding up to the identity operator $\I_A$ ($\I_B$). We denote these measurement operators by $A^x_a$ and $B^y_b$, where $x \in [n_A]$, $a \in [k_A]$, $y \in [n_B]$ and $b \in [k_B]$. Then, the correlation shared by Alice and Bob reads
\begin{equation}\label{eq:correlation}
p_{AB}(a,b | x,y) = \tr[ \rho_{AB} (A^x_a \otimes B^y_b) ],
\end{equation}
specifying the probabilities of observing the outcomes $a$ and $b$, given that the measurements $x$ and $y$ were selected.

The raw data in the protocol corresponds to the pair of strings held by Alice and Bob containing the measurement outcomes and the implemented measurements, collected over all protocol rounds. Since individually Alice and Bob only have access to their marginal statistics, they publicly reveal the measurement settings and outcomes for a fraction of this data to estimate the joint statistics and detect its nonlocality. This part of the dataset is discarded. The secret key is distilled by classically post-processing the remaining dataset with the help of public communication, so that they finally hold identical strings that must appear perfectly random to any third party. 

As stated earlier, in this work we consider what we call \emph{standard} protocols, in which the measurements implemented by Alice and Bob are announced in the key distillation part. Apart from this constraint, the rest of the protocol is arbitrary. This family is quite broad and covers most DIQKD protocols introduced so far \cite{ABG+07,PAB+09,ADF+18,WAP20,HST+20,SBV+20,TLR20}, with only a few exceptions proposed to date (see e.g.~Ref.~\cite{AGM06}).

In order to upper-bound the key rate for a given DIQKD protocol, it suffices to consider a particular model of the eavesdropper, Eve. Here, we restrict the analysis to individual attacks that do not require any quantum memory \cite{PMLA14}. In device-independent protocols, Alice and Bob have no knowledge of the form of the state $\rho_{AB}$ and the measurements $\{A^x_a\}$, $\{B^y_b\}$, and it is precisely this lack of knowledge that Eve makes use of in her attack. In particular, we assume that she knows the precise form of the measurement operators and that she is the one distributing the quantum state (therefore effectively distributing quantum correlations) to Alice and Bob in each round.

In our \textit{convex combination} (CC) attack---originally considered for eavesdroppers limited only by the no-signalling principle \cite{AGM06,AMP06}---Eve distributes local deterministic correlations with certain probabilities that give rise to a local correlation $p^\cL_{AB}(a,b | x,y)$ with overall probability $q_\cL$, and she distributes a nonlocal quantum correlation $p^\cNL_{AB}(a,b|x,y)$ with probability $1-q_\cL$. While presented in this form for the sake of simplicity, Eve can equally implement the attack by fixing the measurements of Alice and Bob and preparing a unique quantum state $\rho_{ABE}$. Eventually, the observed correlation of Alice and Bob takes the form
\begin{equation}\label{eq:bipartite_correlation}
p_{AB}(a,b | x,y) = q_\cL \cdot p^\cL_{AB}(a,b | x,y) + (1-q_\cL) \cdot p^\cNL_{AB}(a,b|x,y),
\end{equation}
and we call $q_\cL \in [0,1]$ the \textit{local weight}. Since nonlocality is necessary for secure DIQKD, in the CC attack Eve maximises $q_\cL$ for the given observed correlation $p_{AB}(a,b | x,y)$ and a judiciously chosen nonlocal quantum correlation $p^\cNL_{AB}(a,b|x,y)$.

We apply the CC attack to the standard DIQKD protocols introduced above. 
Since Alice and Bob announce their inputs for every round, Eve knows their outcomes in all rounds in which she distributes a local correlation. We represent this knowledge by the classical variable $e$, and we write $e = (a,b)$ for the local rounds. On the other hand, we assume in what follows that Eve is not correlated to the nonlocal part of the correlation of Alice and Bob, denoted by $e = ?$. Therefore, for any combination of inputs $x$ and $y$, Alice, Bob and Eve share correlated random variables distributed as
\begin{equation}\label{eq:tripartite}
\begin{split}
p_{ABE}(a,b,e|x,y) & \left. =  q_\cL \cdot p^\cL_{AB}(a,b | x,y) \cdot \delta_{e,(a,b)} \right. \\
& \left. + (1 - q_\cL) \cdot p^\cNL_{AB}(a,b| x, y) \cdot \delta_{e,?} , \right.
\end{split}
\end{equation}
where $\delta$ is the Kronecker delta.

Well-established results in classical cryptography prove that the asymptotic key rate $\rt$ extractable from a dataset of strings distributed according to $p_{ABE}(a,b,e)$ is upper-bounded by the \textit{intrinsic information}~\cite{MW97,CEH+07},
\begin{equation}\label{intrinf}
I(A,B\downarrow E)=\min_{E\rightarrow F} I(A:B | F),
\end{equation}
where $I(A:B | F) = \sum_{f} p_{F}(f) \cdot I(A:B|F=f)$ is the conditional mutual information of $p_{ABF}(a,b,f)$, and the minimisation is taken over all stochastic maps $E\rightarrow F$ that map the variable $E$ (with values $e$) to a new variable $F$ (with values $f$), such that the alphabet size of $F$ is at most that of $E$~\cite{CRW03}. While this minimisation may be hard, any candidate stochastic map provides a valid upper bound. 

When applying this bound to the CC attack, the key rate is upper-bounded by
\begin{equation}\label{eq:conditional_mutual_info}
\rt \leq \sum_{x,y} p_{xy} \cdot I_{xy}(A:B\downarrow E ),
\end{equation}
where the sum runs over all those settings $(x,y)$ from which the key is distilled, $p_{xy}$ is the probability of Alice and Bob choosing the settings $x$ and $y$, respectively, and $I_{xy}(A:B\downarrow E )$ is the intrinsic information of the distribution in Eq.~\eqref{eq:tripartite}. Note that the bound in Eq.~\eqref{eq:conditional_mutual_info} is based only on the observed correlation, without any assumption on the state or the measurements.

\prlparagraph{Nonlocality is not sufficient for DIQKD}
In what follows, we prove that there exist nonlocal correlations that cannot be used for secure key extraction with standard DIQKD. We do this by applying the CC attack on any correlation obtained by performing arbitrary projective measurements on the two-qubit Werner state \cite{Wer89} with \emph{visibility} $v\in [0,1]$,
\begin{equation}\label{eq:Werner}
\rho^v_{AB} = v \ketbraq{\psi_-} + (1-v) \frac{\I}{4},
\end{equation}
where $\ket{\psi_-} = ( \ket{01} - \ket{10} ) / \sqrt{2}$. It is known that for arbitrary (even infinitely many) projective measurements $\{ A^x_a \}$ and $\{ B^y_b \}$, the correlation $p^v_{AB}(a,b|x,y) := \tr[ \rho^v_{AB} (A^x_a \otimes B^y_b) ]$ is local whenever the visibility is at most $v \le v^w_\cL := 999 \cdot 689 \cdot 10^{-6} \cdot \cos^4(\pi / 50) \approx 0.6829$, see Ref.~\cite{HQV+17}. On the other hand, it is also known that there exist projective measurements that give rise to nonlocal correlations for $v \ge v^w_\cNL :\approx 0.6964$, see Ref.~\cite{DBV17}.

Let us consider all DIQKD protocols that use correlations obtained by implementing arbitrarily many projective measurements on the Werner state. The measurements can be written as $A^x_a = \frac12[\I + (-1)^a \va_x \cdot \vec{\sigma}]$ and $B^y_b = \frac12[\I + (-1)^b \vb_y \cdot \vec{\sigma}]$, where $a,b \in \{0,1\}$, $\va_x$ and $\vb_y$ are unit vectors in $\mathbb{R}^3$, and $\vec{\sigma} = (X, Y, Z)$ is a vector containing the Pauli matrices. It is easy to verify that
\begin{equation}\label{eq:Werner_correlation}
\begin{bmatrix}
p^v_{AB}(0,0|x,y) & p^v_{AB}(0,1|x,y) \\
p^v_{AB}(1,0|x,y) & p^v_{AB}(1,1|x,y) 
\end{bmatrix}
= \frac12
\begin{bmatrix}
s^v_{xy} & 1 - s^v_{xy} \\
1 - s^v_{xy} & s^v_{xy}
\end{bmatrix},
\end{equation}
where $s^v_{xy} = \frac12(1 - v \va_x \cdot \vb_y) \in [0,1]$.
%
%

The CC attack we consider is rather intuitive: it uses the nonlocal correlation $p_{AB}^\cNL=p_{AB}^{v = 1}$, and local deterministic correlations that sum up to the correlation for the provable local Werner state, $p_{AB}^\cL=p_{AB}^{v=v_\cL^w}$. It is easy to verify that in this case we have that $q_\cL = q^w_\cL := (1-v)/(1-v_\cL^w)$. For now, let us assume that $s^1_{xy} \equiv s^{v=1}_{xy} \ge \frac12$. This implies that in the ideal ($v=1$) case, the outcomes of Alice and Bob are correlated, i.e., they observe $a=b$ more often than $a \neq b$ [see Eq.~\eqref{eq:Werner_correlation}]. For this reason, in her stochastic relabelling $E \to F$, Eve will attempt to become as correlated to the $a=b$ events as possible, that is, she picks $f = a$ whenever the correlation is local and $a = b$. In order to reduce the conditional mutual information of all the other events, she sets $f = ?$ for all the remaining cases.
The resulting distribution reads
\begin{equation}\label{eq:tripartite_map}
\begin{split}
p_{ABF}(a,b,f|x,y) & \left. =  q^w_\cL \cdot p^{v=v_\cL^w}_{AB}(a,b | x,y) \right. \\
& \left. \times \big[ \delta_{a,b} \cdot \delta_{f,a} + (1 - \delta_{a,b}) \cdot \delta_{f,?} \big] \right. \\
& \left. + (1 - q^w_\cL) \cdot p^{v = 1}_{AB}(a,b| x, y) \cdot \delta_{f,?} . \right.
\end{split}
\end{equation}
Note that a similar distribution can be introduced for the case of $s^1_{xy} < \frac12$, in which case Eve becomes correlated with the $a \neq b$ events. For the distribution in Eq.~\eqref{eq:tripartite_map} we have that
 $I_{xy}(A:B|F=a) =  0$ for all~$a$, so the final bound on the key rate is given by
\begin{equation}\label{eq:keyrate}
\rt \le \sum_{x,y} p_{xy} \cdot p_{F}(?|x,y) \cdot I_{xy}(A:B|F=?),
\end{equation}
where $I_{xy}(A:B|F=f)$ is the mutual information of the distribution $p_{AB|F}(a,b|f,x,y) = p_{ABF}(a,b,f|x,y) / p_{F}(f|x,y)$.

To compute the upper bound, we need to calculate the terms $I_{xy}(A:B|F=?)$ in Eq.~\eqref{eq:keyrate}, that is, the mutual information of the distribution
\begin{widetext}
\begin{equation}
p_{AB|F}(a,b|?,x,y) = \frac{1}{ 2( 1 - q^w_\cL \cdot s_{xy}^w ) }
\begin{bmatrix}
(1 - q^w_\cL) s^1_{xy} & (1 - q^w_\cL) ( 1 - s^1_{xy} ) + q^w_\cL (1 - s_{xy}^w ) \\
(1 - q^w_\cL) ( 1 - s^1_{xy} ) + q^w_\cL (1 - s_{xy}^w ) & (1 - q^w_\cL) s^1_{xy}
\end{bmatrix},
\end{equation}
\end{widetext}
where $s^w_{xy} = s^{v = v^w_\cL}_{xy}$ and $s^1_{xy} = s^{v = 1}_{xy}$. The mutual information is clearly zero whenever $(1 - q^w_\cL) s^1_{xy} = (1 - q^w_\cL) ( 1 - s^1_{xy} ) + q^w_\cL (1 - s_{xy}^w )$, which is equivalent to
\begin{equation}\label{eq:vcrit_s}
v = \frac{ v^w_\cL ( 2 s^1_{xy} - 1 ) + 1 }{ v^w_\cL ( 1 - 2 s^1_{xy} ) + 4 s^1_{xy} - 1} =: v_{xy}.
\end{equation}
Note that while $I_{xy}(A:B|F=?)$ is in general positive for $v < v_{xy}$, a slight modification of the CC attack leads to $I_{xy}(A:B|F=?) = 0$ for any $v \le v_{xy}$.
This is achieved by a stochastic relabelling $E \to F$ in which Eve only maps some fraction $\lambda_{xy}$ of her variables $e = (a,b)$ with $a \neq b$ to $f = ?$, and leaves the remaining fraction $1-\lambda_{xy}$ invariant. It is straightforward to verify that with a properly chosen $\lambda_{xy}$, this relabelling leads to $I_{xy}(A:B|F) = 0$ for any $v \le v_{xy}$.

Also note that $v_{xy}$ is monotonically decreasing in $s^1_{xy}$, and hence, it reaches its lowest possible value at $s^1_{xy} = 1$. This gives rise to the \textit{critical visibility} of the Werner state,
\begin{equation}\label{eq:vcrit_Werner}
v^w_\text{crit} = \frac{ v^w_\cL + 1 }{ 3 - v^w_\cL } \approx 0.7263 > v^w_\cNL \approx 0.6964.
\end{equation}
An analogous derivation yields the same critical visibility for $s^1_{xy} < \frac12$.
From the above arguments, it follows that whenever the visibility is $v^w_\cNL \le v \le v^w_\text{crit}$, Alice and Bob cannot extract a secure key from correlations obtained from the Werner state with any (even infinite) number of projective measurements, even though the state is nonlocal (i.e., there exist projective measurements that, measured on the state, give rise to nonlocal correlations) and distillable \cite{HHH97}. This means that the Bell nonlocality of the observed correlation is in general not sufficient for DIQKD whenever Alice and Bob announce their measurement settings, and, moreover, that there exist nonlocal states that cannot be used for standard DIQKD with projective measurements.

\prlparagraph{CHSH-based protocols}
The most commonly used DIQKD protocols \cite{ABG+07,PAB+09,ADF+18,WAP20,HST+20,SBV+20,TLR20} are based on the maximal violation of the biased CHSH inequality~\cite{AMP12}. In these protocols the shared state is $\rho_{AB} = \ketbraq{ \psi_- }$, and Alice's measurements are described by $\va_0 = (0,0,-1)$ and $\va_1 = (-1,0,0)$, while Bob's measurements are described by $\vb^\theta_0 = (\sin \theta, 0, \cos \theta)$, $\vb^\theta_1 = (-\cos\theta, 0, \sin\theta)$ and $\vb^\theta_2 = (0,0,1)$, where $0 < \theta < \frac{\pi}{2}$. The protocol based on the standard CHSH inequality is reproduced by setting $\theta = \frac{ \pi }{4}$ \cite{ABG+07,PAB+09}. The noisy versions of these protocols with visibility $v$ can be described by sharing a Werner state, and our results from the previous section readily apply. In particular, $v^w_\text{crit}$ in Eq.~\eqref{eq:vcrit_Werner} is a lower bound on the critical visibility for CHSH-based protocols.

However, for a fixed protocol, the bound on the critical visibility can be improved. This is because in the setting of the CHSH-based protocols, the polytope of local correlations is completely characterised \cite{CG04}. One can verify that in the CHSH-based protocols, if Alice and Bob observe a correlation that corresponds to the Werner state with visibility $v$, then this correlation is local if and only if $v \le v^\theta_\cL := 1/(\cos\theta + \sin\theta)$ \footnote{\SupMat}. Therefore, an improved bound on the critical visibility for the CHSH-based protocols is given by
\begin{equation}\label{eq:vcrit_CHSH}
v^\theta_\text{crit} = \frac{ v^\theta_\cL + 1 }{ 3 - v^\theta_\cL } > v^\theta_\cL \quad \forall \theta \in (0,  \pi/2).
\end{equation}
That is, for a range of visibilities for which the observed correlation is nonlocal, Alice and Bob cannot extract a secure key. Note that the same critical visibility holds for the recently introduced modification of the standard CHSH-based protocol in Ref.~\cite{SGP+20}, where the authors add a fourth setting for Bob. Indeed, since the local polytope is completely characterised in this case as well \cite{CG04}, one can verify that the correlation becomes local at the same visibility $v^\theta_\cL$ \cite{Note1}.

\begin{figure}[t!]
\centering
\includegraphics[width=\columnwidth]{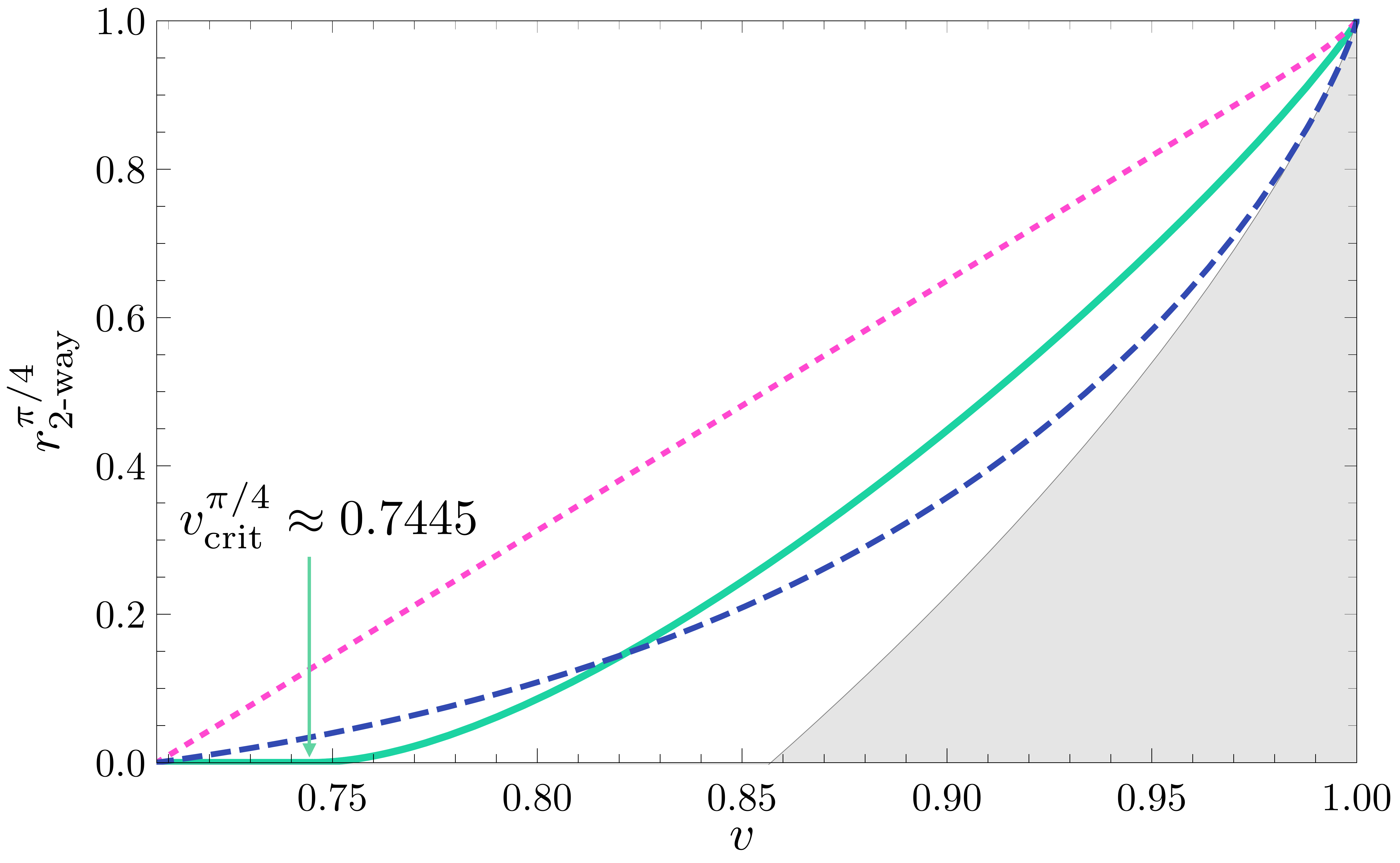}
\caption{Upper bounds on the two-way key rate for the standard CHSH protocol in terms of the visibility. The dotted line is the upper bound from \cite{KWW20}, the dashed line is the upper bound from \cite{AL20}, and the solid line is the bound in Eq.~\eqref{eq:twowaybound}. Note that the visibility can be converted into the CHSH violation $S$ via $S = 2 \sqrt{2} v$. The shaded area represents the lower bound from \cite{ABG+07}.
\label{fig:twoway}}
\end{figure}

Last, we note that in the CHSH-based protocols, Alice and Bob usually extract their key from the setting pair $x=0$ and $y=2$, by setting $p_{02}$ in Eq.~\eqref{eq:conditional_mutual_info} arbitrarily close to 1. In this case, it is possible to compute the upper bound in Eq.~\eqref{eq:keyrate} for any visibility $v \ge v^\theta_\text{crit}$, and we get
%
%
\begin{equation}\label{eq:twowaybound}
\begin{split}
\rt^\theta(v) & \left. \le
2(1 - s^\theta q^\theta_\cL ) + (1 - q^\theta_\cL) \log_2\!\left[ \frac{1 - q^\theta_\cL}{2(1 - s^\theta q^\theta_\cL )}\right] \right. \\
& \left. + q^\theta_\cL(1 - s^\theta) \log_2\! \left[ \frac{q^\theta_\cL(1 - s^\theta)}{2(1 - s^\theta q^\theta_\cL )}\right] \right.
\end{split}
\end{equation}
where $s^\theta = \frac12(1 + v^\theta_\cL)$ and $q^\theta_\cL = (1-v)/(1-v^\theta_\cL)$.
In Fig.~\ref{fig:twoway} we plot the bound for the standard CHSH protocol ($\theta = \frac{ \pi }{4}$), and show that it outperforms the recently derived upper bounds \cite{KWW20,AL20} near the critical visibility. In the Supplemental Material we also describe a two-dimensional region in the set of quantum correlations corresponding to correlations from the biased CHSH protocol that are nonlocal but cannot be used to extract a key using standard DIQKD \cite{Note1}.

\prlparagraph{Discussion}
We introduced a generic tool for upper-bounding DIQKD key rates using a simple eavesdropping attack. Using our tool, we showed that Bell nonlocality is not sufficient for secure DIQKD when the honest parties announce their measurement settings. Our results also imply that all the commonly used DIQKD protocols 
become insecure in the noisy case already in the nonlocal regime, even when assisted by arbitrary two-way communication. Our analysis does not prove that the considered nonlocal correlations are useless for secure key distribution, but it shows that the standard reconciliation---where the settings are announced by both parties---does not work for all nonlocal correlations.

Given the above, one possibility to lower the stringent requirements on noise parameters is to employ protocols such as that of Ref.~\cite{AGM06}, in which only one party announces their settings. Indeed, for this protocol we were not able to find an upper bound that vanishes in the nonlocal regime. Whether a secure key can be distilled from all nonlocal correlations using these protocols is an open question that deserves further investigation. Another possibility for improving the key rates extractable from a given quantum state is to employ measurements that are not projective. However, we note that no state is known thus far that is local for all arrangements of projective measurements, while exhibiting nonlocality for some arrangement of non-projective measurements. Hence, the critical visibilities of the Werner state derived in this work also hold for all the hitherto studied arrangements of non-projective measurements. Another question worth investigating is whether tighter upper bounds can be derived using collective or coherent attacks. Nonetheless, let us note that our (individual) CC attack can be applied to a broad class of DIQKD protocols, and gives rise to bounds on the critical visibility in experimentally relevant scenarios. We elaborate on these findings in \footnote{In preparation}.

\prlparagraph{Acknowledgements}
We thank Erik Woodhead, Stefano Pironio, J\k{e}drzej Kaniewski, Karol Horodecki, Filip Rozp\k{e}dek and Goh Koon Tong for fruitful discussions, and Marco T\'ulio Quintino for pointing out Ref.~\cite{DBV17}.
We acknowledge support from the Government of Spain (FIS2020-TRANQI and Severo Ochoa CEX2019-000910-S), Fundaci\'o Cellex, Fundaci\'o Mir-Puig, Generalitat de Catalunya (CERCA, AGAUR SGR 1381 and QuantumCAT), the ERC AdG CERQUTE, the AXA Chair in Quantum Information Science, the EU Quantum Flagship project QRANGE, and the Foundation for Polish Science within the “Quantum Optical Technologies” project carried out within the International Research Agendas programme cofinanced by the European Union under the European Regional Development Fund. MBJ acknowledges funding from the European Union's Horizon 2020 research and innovation programme under the Marie Sk\l{}odowska-Curie grant agreement No 847517.


%

\newpage

\onecolumngrid

\appendix

\vspace{1cm}

\begin{center}
\large{\bf Supplemental Material}
\end{center}

\section{Local visibility for CHSH-based protocols}\label{sec:local}

Consider the correlation $p^{\theta,v}_{AB}(a,b|x,y) = \tr\{ \rho^v_{AB}[A^x_a \otimes B^y_b(\theta)]\}$, obtained by measuring the Werner state
\begin{equation}\label{eqs:Werner}
\rho^v_{AB} = v \ketbraq{\psi_-} + (1-v)\frac{\I}{4}
\end{equation}
with the projective measurements $A^x_a = \frac12[\I + (-1)^a \va_x \cdot \vec{\sigma}]$ and $B^y_b(\theta) = \frac12[\I + (-1)^b \vb^\theta_y \cdot \vec{\sigma}]$ with $a,b \in \{0,1\}$, given by
\begin{equation}\label{eqs:CHSH_mmt}
\begin{split}
\va_0 = (0,0,-1), \quad  \va_1 = (-1, & \left. 0,0) \right. \\
\vb_0^\theta = (\sin\theta, 0, \cos\theta), \quad \vb_1^\theta = (-\cos\theta, 0, & \left. \sin\theta), \quad \vb_2^\theta = (0,0,1). \right.
\end{split}
\end{equation}
Note that with the above notation, we have that $p^{\theta,v}_{AB}(a,b|x,y) = \frac14[ 1 - v(-1)^{a+b} \va_x \cdot \vb_y^\theta ]$.

In this section, we show that $p^{\theta,v}_{AB}(a,b|x,y)$ is local if and only if
\begin{equation}
v \le v^\theta_\cL = \frac{1}{ \cos\theta + \sin\theta }.
\end{equation}

In order to prove this, we note that for the case of two binary measurements on Alice's side and three binary measurements on Bob's side, the polytope of local correlations is completely characterised [28]. In particular, all the facets that correspond to non-trivial constraints (i.e., do not correspond to the positivity and normalisation of the probabilities) are of the CHSH-type:
\begin{equation}\label{eqs:CHSH}
-2 \le \langle A_{x_0} B_{y_0} \rangle + \langle A_{x_0} B_{y_1} \rangle + \langle A_{x_1} B_{y_0} \rangle - \langle A_{x_1} B_{y_1} \rangle \le 2 \quad x_0,x_1 \in \{0,1\}, \, y_0, y_1 \in \{0,1,2\}, x_0 \neq x_1, y_0 \neq y_1,
\end{equation}
where
\begin{equation}\label{eqs:correlator}
\langle A_x B_y \rangle = p_{AB}(0,0|x,y) + p_{AB}(1,1|x,y) - p_{AB}(0,1|x,y) - p_{AB}(1,0|x,y)
\end{equation}
are the \textit{correlators}. In other words, a correlation $p_{AB}(a,b|x,y)$ in this setting is local if and only if it satisfies all the inequalities in Eq.~\eqref{eqs:CHSH}.

Let us denote the correlators of $p^{\theta,v}_{AB}(a,b|x,y)$ by $\langle A_x B_y \rangle^{\theta,v}$. Note that $\langle A_x B_y \rangle^{\theta,v} = -v \va_x \cdot \vb_y^\theta$, and hence $\langle A_x B_y \rangle^{\theta,v} = v \cdot \langle A_x B_y \rangle^{\theta,1}$. Therefore, the statement that $p^{\theta,v}_{AB}(a,b|x,y)$ is local is equivalent to 
\begin{equation}\label{eqs:CHSH_v}
-2 \le v \cdot \mathcal{S}^\theta_{x_0, x_1, y_0, y_1} \le 2 \quad \forall x_0,x_1 \in \{0,1\}, \, y_0, y_1 \in \{0,1,2\}, x_0 \neq x_1, y_0 \neq y_1,
\end{equation}
where we have defined
\begin{equation}
\mathcal{S}^\theta_{x_0, x_1, y_0, y_1} = \langle A_{x_0} B_{y_0} \rangle^{\theta,1} + \langle A_{x_0} B_{y_1} \rangle^{\theta,1} + \langle A_{x_1} B_{y_0} \rangle^{\theta,1} - \langle A_{x_1} B_{y_1} \rangle^{\theta,1}.
\end{equation}
A straightforward computation yields
\begin{equation}\label{eqs:CHSH_perms}
\begin{split}
\mathcal{S}^\theta_{0,1,0,1} &=  2(\cos\theta+\sin\theta)   \\
\mathcal{S}^\theta_{0,1,0,2} &=  1+\cos\theta+\sin\theta   \\
\mathcal{S}^\theta_{0,1,1,0} &=   -\cos\theta-\sin\theta  \\
\mathcal{S}^\theta_{0,1,1,2} &=  1-\cos\theta+\sin\theta  \\
\mathcal{S}^\theta_{0,1,2,0} &=   1+\cos\theta-\sin\theta  \\
\mathcal{S}^\theta_{0,1,2,1} &=   1+\cos\theta+\sin\theta  \\
\mathcal{S}^\theta_{1,0,0,1} &=   0  \\
\mathcal{S}^\theta_{1,0,0,2} &=   -1+\cos\theta+\sin\theta  \\
\mathcal{S}^\theta_{1,0,1,0} &=   2(-\cos\theta+\sin\theta)    \\
\mathcal{S}^\theta_{1,0,1,2} &=   -1-\cos\theta+\sin\theta  \\
\mathcal{S}^\theta_{1,0,2,0} &=   1-\cos\theta+\sin\theta  \\
\mathcal{S}^\theta_{1,0,2,1} &=   1-\cos\theta-\sin\theta . 
\end{split}
\end{equation}
By comparing Eqs.~\eqref{eqs:CHSH_v} and \eqref{eqs:CHSH_perms}, it is straightforward to verify that $p^{\theta,v}_{AB}(a,b|x,y)$ is local if and only if $v \le v^\theta_\cL = 1/(\cos\theta + \sin\theta)$.

This argument naturally extends to the CHSH-based protocol recently introduced in Ref.~[30]. This protocol is based on the state in Eq.~\eqref{eqs:Werner} and the measurements in Eq.~\eqref{eqs:CHSH_mmt} (with $\theta = \frac{\pi}{4}$, but our argument works for arbitrary $\theta$), with an added setting for Bob, described by $\vb^\theta_3 = (1,0,0)$. These new correlations trivially satisfy Eq.~\eqref{eqs:CHSH_perms}, and additionally, it is easy to verify that
\begin{equation}
\mathcal{S}^\theta_{x_0,x_1,y,3} = \mathcal{S}^\theta_{x_1,x_0,y,2} \quad \text{and} \quad
\mathcal{S}^\theta_{x_0,x_1,3,y} = \mathcal{S}^\theta_{x_0,x_1,2,y}.
\end{equation}

The local polytope is completely characterised in this scenario as well [28], and all the non-trivial facets are of the CHSH-type:
\begin{equation}\label{eqs:CHSH_4}
-2 \le \langle A_{x_0} B_{y_0} \rangle + \langle A_{x_0} B_{y_1} \rangle + \langle A_{x_1} B_{y_0} \rangle - \langle A_{x_1} B_{y_1} \rangle \le 2 \quad x_0,x_1 \in \{0,1\}, \, y_0, y_1 \in \{0,1,2,3\}, x_0 \neq x_1, y_0 \neq y_1.
\end{equation}
Hence, it is clear that these new correlations are also local in the noisy case if and only if $v \le v^\theta_\cL$.

\section{A two-dimensional region of quantum correlations with zero key}

\begin{figure}[t!]
\centering
\includegraphics[width=0.4\columnwidth]{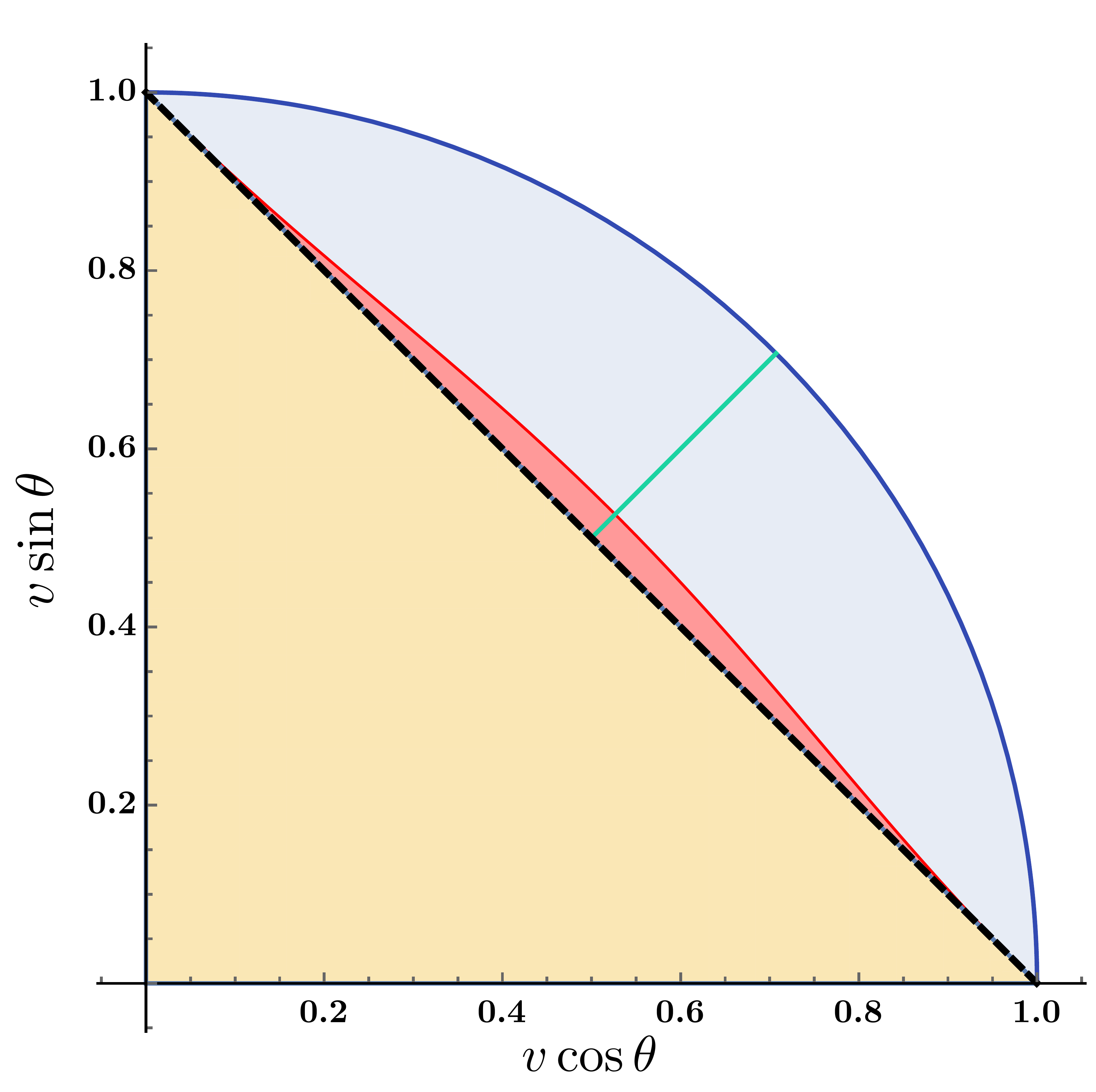}
\caption{We depict a 2-dimensional slice of the set of correlations in the scenario with two binary measurements on Alice's side and three binary measurements on Bob's side. We can see that there are three different regions: in red, we show the region of nonlocal quantum correlations that cannot be used for key extraction if the honest parties announce their settings. In yellow, we depict (part of) the region of local correlations, and the dashed black line represents the CHSH facet of the local polytope. In light blue, we show the region of nonlocal quantum correlations for which our upper bound on the key rate is non-zero. The dark blue curve corresponds to the maximal possible quantum violation of the biased CHSH inequalities, that is, the ideal correlations for $0<\theta<\pi/2$. Finally, the green line corresponds to the correlations obtained from the noisy standard CHSH protocol.}\label{fig:inset}
\end{figure}

In this section, we describe a two-dimensional region in the set of quantum correlations that is nonlocal but cannot be used to extract a secure key using standard DIQKD. This region corresponds to correlations based on the biased CHSH inequality (see the main text), and is depicted in Fig.~\ref{fig:inset}. The points in the figure are based on correlations with two inputs and two outputs, $p_{AB}(a,b|x,y) = \tr[ \rho_{AB} (A^x_a \otimes B^y_b)]$, such that $a,b,x,y \in \{0,1\}$. The full set of these correlations can be embedded in an 8-dimensional real vector space [28]. To see this, let us define the \textit{observables}
\begin{equation}
A_x \equiv A^x_0 - A^x_1 \quad \text{and} \quad B_y \equiv B^y_0 - B^y_1.
\end{equation}
The original measurement operators can be recovered from the observables via
\begin{equation}
A^x_a = \frac12[ \I + (-1)^a A_x ] \quad \text{and} \quad B^y_b = \frac12[ \I + (-1)^b B_y ].
\end{equation}
In terms of these observables, we define the \textit{marginals}
\begin{equation}
\langle A_x \rangle = \tr[ \rho_{AB} (A_x \otimes \I) ] \quad \text{and} \quad \langle B_y \rangle = \tr[ \rho_{AB} (\I \otimes B_y) ]
\end{equation}
and the \textit{correlators}
\begin{equation}
\langle A_x B_y \rangle = \tr [ \rho_{AB} (A_x \otimes B_y) ]
\end{equation}
[notice that these correlators are the same as those in Eq.~\eqref{eqs:correlator}]. Indeed, any correlation can be written in terms of the marginals and the correlators as
\begin{equation}
p_{AB}(a,b|x,y) = \tr[ \rho_{AB} (A^x_a \otimes B^y_b)] = \frac14 \big[ 1 + (-1)^a \langle A_x \rangle + (-1)^b \langle B_y \rangle + (-1)^{a+b} \langle A_x B_y \rangle \big].
\end{equation}

In order to describe a 2-dimensional slice of the set of correlations in this scenario, we need to describe all correlators and marginals in terms of affine functions of two real variables, which we will denote by $s$ and $t$. The slice we are interested in is given by 
\begin{equation}\label{eqs:slice}
\begin{split}
& \left. \la A_x \ra = \la B_y \ra = 0 \quad \forall x,y \right. \\
& \left. \la A_0 B_0 \ra = s \right. \\
& \left. \la A_0 B_1 \ra = t \right. \\
& \left. \la A_1 B_0 \ra = t \right. \\
& \left. \la A_1 B_1 \ra = -s. \right.
\end{split}
\end{equation}
The set of quantum correlations in this case is bounded by the relation $s^2 + t^2 \le 1$, which corresponds to the maximal possible quantum violation of the biased CHSH inequalities [10]. The local set is bounded by four constraints, $\pm s \pm t \le 1$, which are various relabellings of the CHSH inequality, and the inequality $s + t \le 1$ corresponds to the standard CHSH inequality
\begin{equation}
\la A_0 B_0 \ra + \la A_0 B_1 \ra + \la A_1 B_0 \ra - \la A_1 B_1 \ra \le 2.
\end{equation}

Using this parametrisation, the local points on the CHSH facet correspond to $s + t = 1$, with the two endpoints being $(s, t) = (1, 0)$ and $(s, t) = (0, 1)$. The boundary of the quantum set (with all points being extremal) corresponds to the curve $s^2 + t^2 = 1$. We will be interested in the region $s + t > 1$, i.e., correlations violating the standard CHSH inequality, while still being quantum, i.e., $s^2 + t^2 \le 1$. It is convenient to parametrise this region with the polar coordinates $(s,t) = (v \cos \theta, v \sin\theta)$:
\begin{equation}\label{eqs:inset_2222}
\begin{split}
& \left. \la A_x \ra = \la B_y \ra = 0 \quad \forall x,y \right. \\
& \left. \la A_0 B_0 \ra = v \cos\theta \right. \\
& \left. \la A_0 B_1 \ra = v \sin\theta \right. \\
& \left. \la A_1 B_0 \ra = v \sin\theta \right. \\
& \left. \la A_1 B_1 \ra = -v \cos\theta, \right.
\end{split}
\end{equation}
where $0 < \theta < \frac{\pi}{2}$ and $ v^\theta_\cL < v \le 1 $ [let us recall that $v^\theta_\cL = 1/(\cos\theta + \sin\theta)$, see the previous section]. 

The region depicted in Fig.~\ref{fig:inset} corresponds to DIQKD protocols with two binary measurements on Alice's side and \textit{three} binary measurements on Bob's side. The correlation observed in the protocol is given by Eq.~\eqref{eqs:inset_2222} with an added binary measurement, $y=2$, on Bob's side, that gives rise to the final correlation
\begin{equation}\label{eqs:inset_2223}
\begin{split}
& \left. \la A_x \ra = \la B_y \ra = 0 \quad \forall x,y \right. \\
& \left. \la A_0 B_0 \ra = v \cos\theta \right. \\
& \left. \la A_0 B_1 \ra = v \sin\theta \right. \\
& \left. \la A_0 B_2 \ra = v \right. \\
& \left. \la A_1 B_0 \ra = v \sin\theta \right. \\
& \left. \la A_1 B_1 \ra = -v \cos\theta, \right. \\
& \left. \la A_1 B_2 \ra = 0. \right. 
\end{split}
\end{equation}
It is clear that this correlation can be obtained by measuring the state in Eq.~\eqref{eqs:Werner} with the projective measurements described in Eq.~\eqref{eqs:CHSH_mmt}.

The region depicted in Fig.~\ref{fig:inset} corresponds to the set of correlations of the form \eqref{eqs:inset_2223} with $0 < \theta < \frac{\pi}{2}$ and $v^\theta_\cL < v \le 1$ [this region is depicted using the coordinates $(s,t)=(v\cos\theta,v\sin\theta)$]. The black dashed line corresponds to $v = v^\theta_\cL$, and the blue curved boundary on the edge corresponds to $v = 1$. The green line is given by $\theta = \frac{\pi}{4}$. The red region corresponds to $v^\theta_\cL < v \le v^\theta_\text{crit} = (v^\theta_\cL + 1) / (3 - v^\theta_\cL)$. As explained in the main text, Alice and Bob cannot extract a secure key if they observe a correlation in the red region and publicly announce their settings for each round, even though these correlations are nonlocal.

\end{document}